\documentclass{PoS}

\usepackage{graphicx}
\usepackage{amsmath}
\usepackage{amssymb}
\usepackage{float}
\usepackage{xspace}
\usepackage{url}
\usepackage{booktabs}
\usepackage[utf8]{inputenc}
\usepackage[english]{babel}
\usepackage[autostyle, english = american]{csquotes}
\usepackage[sort, numbers]{natbib}
\graphicspath{{images/}}

\newcommand{\jpsi}{\ensuremath{J/\psi}\xspace}
\newcommand{\pc}{\ensuremath{P_c(4450)}\xspace}

\title{Heavy Quarkonium Production at Threshold: from JLab to EIC}

\ShortTitle{Heavy Quarkonium Production from JLab to EIC}

\author{S. Joosten\\
        Temple University\\
        E-mail: \email{sylvester.joosten@temple.edu}}

\author{\speaker{Z.-E. Meziani}\\
        Temple University\\
        E-mail: \email{meziani@temple.edu}}

\abstract{In this contribution we present opportunities to address questions about the origin of mass and spin, probe the existence and nature of the LHCb pentaquark state, and probe the color Van der Waal forces among two color neutral hadrons. The key reaction is elastic production of heavy quarkonia ($J/\psi$ and $\Upsilon$) on the nucleon from threshold to large nucleon-quarkonium invariant masses. This is possible when combining the energy range of two high luminosity facilities, Jefferson Lab 12 GeV and an electron ion collider (EIC).}

\FullConference{QCD Evolution 2017\\
		22-26 May, 2017\\
		Jefferson Lab Newport News, VA - USA}

\begin{document}

\section{Introduction}
It is well known that a mass inventory of the visible universe is commensurate with budgeting for the mass of the nuclei at the heart of atoms.
The mass of each atomic nucleus is a result of the mass of its constituent protons and neutrons and their binding energy. 
However, at a more fundamental level, the mass of the simplest nucleus, the proton sitting at the heart of the hydrogen atom, is mostly born out of the energy stored in its massless gluons, its almost massless quarks and their color interactions, rather than the intrinsic mass of the up and down quarks that roam its confined volume.
This makes grasping the fundamental structure of the nucleon, and hadrons in general, critical to our understanding of the visible mass of the universe.
Today, it is accepted that Quantum Chromodynamics (QCD), the gauge theory of strong interactions, plays a central role in our understanding of nucleon structure.
Indeed, this theory provides the means to interpret high energy scattering processes among the constituent particles of hadrons (quarks and gluons) using perturbative QCD, taking advantage of factorization theorems and evolution equations similar to QED.
At the same time QCD offers a path to unravel the non-perturbative structure of hadrons using lattice QCD, a powerful {\it ab initio} numerical method. 

Here we are interested in the experimental exploration of the origin of the proton mass and spin, 
the size of the Van der Waals force in QCD,
and the existence and nature of the LHCb pentaquark 
through exclusive measurements of the elastic electro- and photo-production of $J/\psi$ and $\Upsilon$ at the 12 GeV upgrade of Jefferson Lab and a possible future electron ion collider (EIC).

\subsection{Origin of the proton spin and mass}
Today we are at the dawn of exploring the contribution of the orbital angular momentum of the valence quarks at the 12 GeV upgraded Jefferson Lab. 
Polarized deep inelastic scattering (DIS) experiments at CERN, SLAC, DESY, and Jefferson Lab and polarized proton-proton collisions experiments at BNL have shaped our present knowledge of the net contribution of the helicity of quarks to the nucleon spin, which is about 30\%, and that of gluons, which is found to be positive amid a large uncertainty.
These experimental facts are consistent with the latest lattice QCD calculations performed at the physical pion mass~\cite{Alexandrou:2017oeh}. 
The latter show a contribution of the helicity of quarks consistent with the experimental results but, in addition, provide quantitative results for the quarks orbital angular momentum and the total angular momentum contribution of gluons.
The experimental quest for the contribution of the orbital angular momentum of the quarks is about to start with the upgrade of Jefferson Lab to 12 GeV~\cite{Dudek:2012vr} while for the total angular momentum of  gluons one has to wait for an electron ion collider~\cite{Accardi:2012qut}.   

While the origin of the nucleon spin in terms of its constituents partons (quarks and gluons) has for the last thirty years taken center stage with measurements of the spin structure functions in lepton deep inelastic scattering (DIS) off the nucleon, the origin of the nucleon mass has received little attention on the experimental side.
This difference might find its origin in the fact that, while the proton spin lend itself to a frame independent decomposition~\cite{Ji:1996ek}, this is is not the case for its mass decomposition.
Nevertheless, it is natural to discuss the mass of the nucleon in its rest frame as originally noted by Ji~\cite{Ji:1995sv} in its mass decomposition and more recently by Lorc\'e~\cite{Lorce:2017xzd}.

The use of {\it ab initio} lattice QCD to evaluate the mass of hadrons has proven to be very successful~\cite{Durr:2008zz,Aoki:2008sm,Aoki:2009ix} and a testament that, undoubtedly, QCD is the right theory to describe the non-perturbative aspects of nucleon structure. However, the origin of the nucleon mass in term of its constituents has yet to be articulated in lattice QCD. Exploring and unraveling the nucleon mass in lattice QCD or in QCD model based approaches~\cite{Roberts:2016vyn} is a worthwhile goal that enable us to better understand how the total mass of the nucleon emerges from the constituents, similar to the case of atomic or nuclear physics. This goal is achieved by a mass decomposition in terms of the masses and energy carried by the constituents in the rest frame of the nucleon using the QCD energy momentum tensor. Starting from scale invariance and using the QCD lagrangian, the trace of energy momentum tensor $\Theta^{\mu}_{\mu}$ is given by the well known result~\cite{Peskin:1995ev}:
\begin{equation}
\Theta^{\mu}_{\mu} = \frac{\beta(g)}{2g}G^{a}_{~\rho\sigma}G^{a~\rho\sigma} + \sum_{f}(1+\gamma_m)m_f\bar{\psi}\psi,
\end{equation}
where $g$ in the strong coupling, and $m_f$ is the quark mass for quark flavor $f$.
From Poincar\'e invariance of a spinless proton, we have 
\begin{equation}
\langle P'\vert \Theta^{\mu\nu} \vert P \rangle = (P'^{\mu}P^{\nu}+ P^{\mu}P'^{\nu})A(Q^2),
\end{equation}
with $P'= P+q$ and $Q^2 = -q^2$ and $A(Q^2)$ the gravitational form factor with the momentum sum rule $A(Q^2=0) = 1$. For $P=P'$ and in the chiral limit where the quarks are massless, we have
\begin{equation}
M^2 = \frac{\beta(g)}{4g}\langle P \vert G^2 \vert P\rangle.
\end{equation}

In the above expression it is clear that most of the mass of the nucleon would originate from the dynamics of the quantum theory, in this case QCD.
Two examples of mass decomposition in the rest frame of the nucleon are given by Ji~\cite{Ji:1995sv} and more recently by Lorc\'e~\cite{Lorce:2017xzd}.
These decompositions, while not unique, are very useful to speak about the nucleon mass in terms of its constituents, similar to other fields of Physics.
Taking the example of the mass separation in terms of quarks and gluons kinetic and potential energies, quark masses and the trace anomaly in the rest frame of the proton we write~\cite{Ji:1995sv}:
\begin{equation}
M = \left.\frac{\langle P\vert H_{QCD}\vert P \rangle}{\langle P\vert P\rangle } \right|_{\text{rest frame}}.
\end{equation}
$H_{QCD}$ is the Hamiltonian operator partitioned following $H_{QCD}=H_q+H_g+H_m+H_a$ where $H_q$ is the quarks kinetic and potential energy, $H_g$ is the gluons kinetic and potential energy, $H_m$ is the quarks mass contribution and $H_a$ is the trace anomaly, all given consecutively by:

\begin{eqnarray}
H_q &=& \int d^3\vec x \psi^{\dagger}\left ( -i\mathbf {D} \cdot \alpha \right )\psi, \\
H_g &=& \int d^3\vec x \frac{1}{2}\left ( E^2 +B^2 \right ),\\
H_m &=& \int d^3\vec x \frac{1}{4} \left ( 1+\gamma_m \right ) \bar \psi m \psi, \\
H_a &=&\int d^3\vec x \frac{1}{4}\beta(g_s)\left ( \mathbf{E}^2 - \mathbf{B}^2 \right ).
\end{eqnarray}
Here $\mathbf{E}$ and $\mathbf{B}$ are the chromo-electric and -magnetic color fields respectively.
Taking the expectation value of each term in the rest frame of the proton lead to the following mass decomposition
\begin{equation}
M= M_q + M_g + M_m + M_a,
\end{equation}
where each mass term is expressed as a fraction of the nucleon mass as:
\begin{eqnarray}
M_q &=& \frac{3}{4} \left (a-\frac{b}{1+\gamma_m} \right ) M, \\
M_g &=& \frac{3}{4} \left ( 1-a \right ) M, \\
M_m &=& \frac{4+\gamma_m}{4(1+\gamma_m )}b M,\\
M_a &=& \frac{1}{4}\left ( 1-b \right ) M,
\end{eqnarray}
$a$ and $b$ can be evaluated using lattice QCD or extracted from experimental measurements. 
Similar to the case of the spin sum rule of the nucleon, to complete the proton mass decomposition 
every term in the decomposition needs to be experimentally accessible.
The trace anomaly term $M_a$ is the only component in this decomposition that has not been determined experimentally, nor by a lattice calculation.
The trace anomaly involves a matrix element of the difference of the squares between the color electric and magnetic fields.
However, it was suggested that $M_a$ can be accessed through quarkonium production on a proton close to threshold~\cite{Kharzeev:1995ij,Kharzeev:1998bz,Gryniuk:2016mpk}.
Due to technical difficulties in dealing with gluons, a direct lattice calculation of the trace anomaly has not been performed yet, there is hope that this situation would change in the next few years in time for comparison with the experiments.

\subsection{LHCb charmed pentaquark}
After the recent discovery of the LHCb pentaquarks~\cite{Aaij:2015tga}, many theoretical papers~\cite{Liu:2015fea,Karliner:2015ina,Chen:2015loa,Eides:2015dtr,Wang:2015jsa,Karliner:2015voa,Kubarovsky:2015aaa,Guo:2015umn,Blin:2016dlf} followed shortly after to explain the possible existence and nature of such states.
However, without further independent measurements, it is unclear whether the formed exotic states can be unambiguously identified as resonances.
If the $P_c(4380)$ and $P_c(4450)$ are truly new states, they can take the form of a five-quark state, 
or a molecular bound state, e.g., between charmonium $(2S)$ and the proton~\cite{Eides:2015dtr}, or $\Sigma_c$ and $\bar D^*~$\cite{Lu:2016nnt,Huang:2016tcr}.
On the other hand, it is also possible that the observed states are a purely kinematic effect, such as a consequence of kinematic enhancements through the anomalous triangle singularity (ATS)~\cite{Liu:2015fea}, or other final state interactions that lead to a threshold enhancement~\cite{Bai:2003sw}.
In Hall C we devised an experiment~\cite{Meziani:2016lhg} that, in principle, will put to rest many of the interpretations of the high mass ($P_c= 4450$ MeV) LHCb state.
The observation of the $P_c(4450)$ state in direct photo-production would provide strong evidence of its resonant nature.

\section{12 GeV $J/\psi$ experiments at JLab in hall A and hall C}
In what follows we shall describe approved measurements planned at Jefferson Lab in Hall A and Hall C to measure the $J/\psi$ electro- and photo-production cross sections at threshold and address the determination of the parameter $b$ discussed in the mass decomposition above but never measured before nor directly calculated using lattice QCD. This parameter requires that one determines the real part of the photo-production amplitude at threshold of this exclusive reaction. While we know that the real part of the production amplitude dominates the threshold region it would be beneficial to verify it experimentally through an asymmetry measurement of the interference between Bethe-Heitler and $J/\psi$ production~\cite{Gryniuk:2016mpk} and determine independently the parameter $b$ from a cross section measurement and an asymmetry measurement close to threshold.

\subsection{A measurement of $J/\psi$ electro and photo-production at threshold using SoLID in Hall A \label{jlab} }

In Hall A, a new project known as the solenoidal large intensity device (SoLID) is under development~\cite{Chen:2014psa}. One of the experiments that will address the proton mass problem is described in detail in Ref.~\cite{SoLIDjpsi:proposal}, and was approved by the Jefferson Lab program advisory committee in 2012. 

In this experiment (known as JLab experiment E12-12-006) a 3$\mu$A electron beam of about 11 GeV incident on a 15 cm hydrogen target will provide for a total integrated luminosity of 43.2 ab$^{-1}$. Using the SoLID large acceptance spectrometer all particles in the final state will be detected for an absolute calibration, namely the scattered electron, the $J/\psi\rightarrow e^+e^-$ decay pair  and the proton. However, for a high statistics data collection in photo-production we will detect the $e^+e^-$ decay pair and the recoiling proton at a rate of 1627 $J/\psi$  events per day, while in electro-production we will use a 3-fold coincidence between the scattered electron and the $e^+e^-$ decay pair at a rate of 86 $J/\psi$ events per day. This approved experiment will collect 50 days of data with 10 days of calibration. It will provide for the highest statistics in the threshold region among the $J/\psi$ elastic production experiments planned at Jefferson Lab.

\begin{figure}
\centering
\includegraphics[width=.38\textheight]{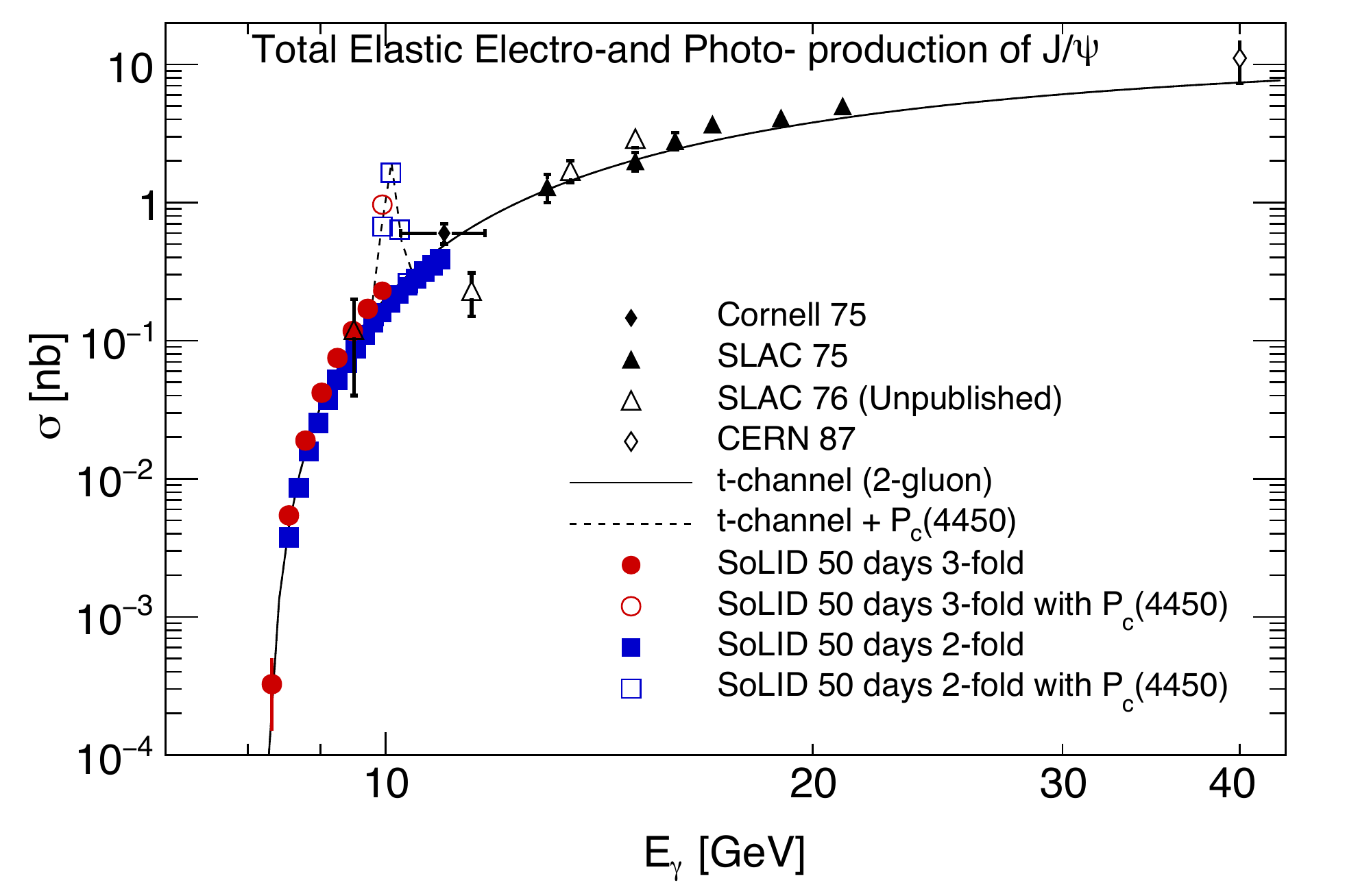}
\includegraphics[width=.265\textheight]{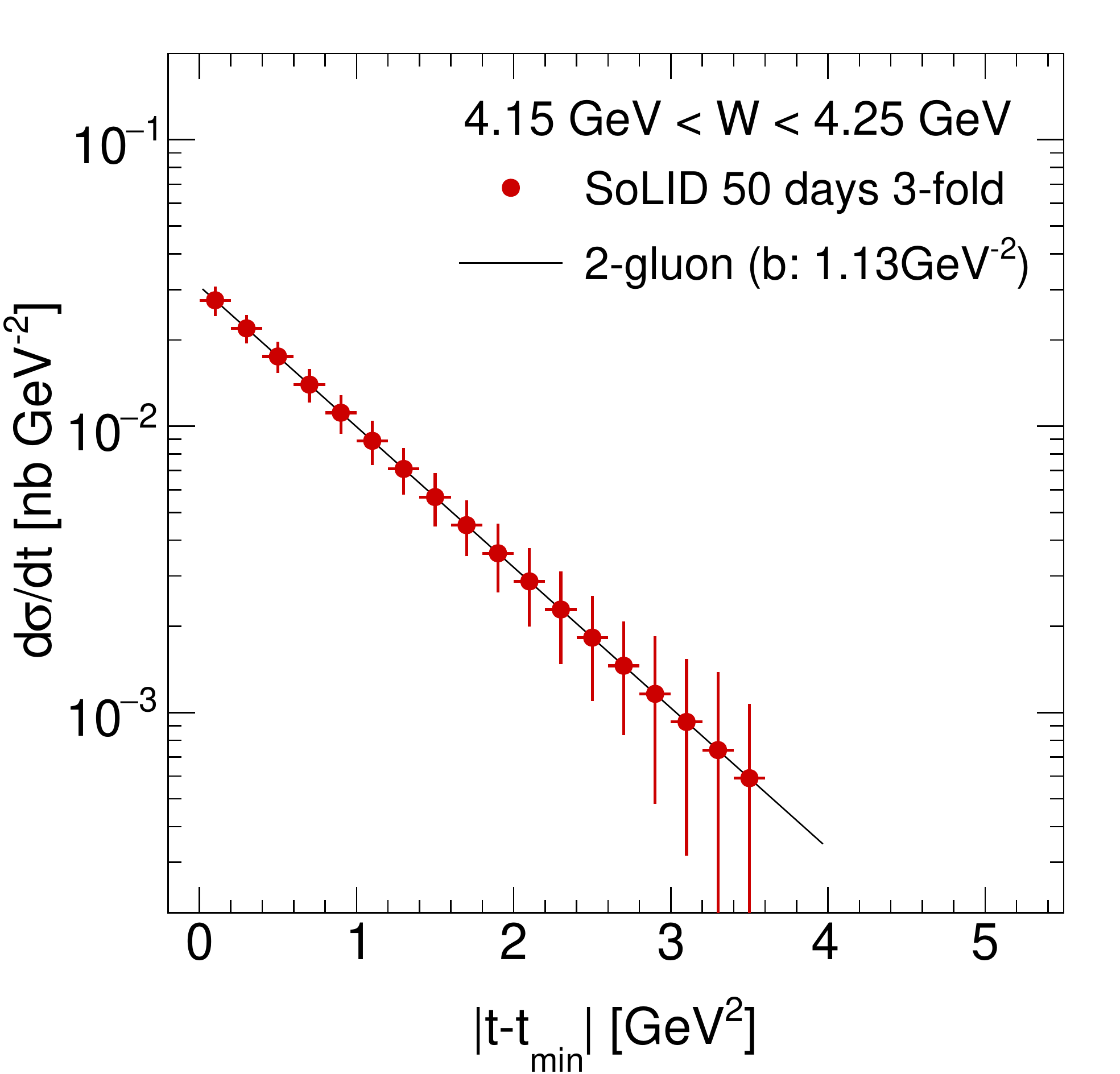}
\caption{
Left: Projected uncertainties of the total elastic $J/\psi$ electro (filled circles) and photo (filled squares)-production cross section using the SoLID detector as a function of photon energy $E_\gamma$. 
The projections are based on the 2-gluon exchange model~\cite{Brodsky:2000zc}.
To enhance the visual representation, the central values of our projections are positioned at 1.2 (0.8) times the predicted total cross section for electro-production (photo-production).
Also plotted (open circles and squares) is the contribution of the larger mass LHCb pentaquark.
Right: differential cross section as a function of $ \vert t-t_{min}\vert$ in the case of electro-production in a $W$ bin very close to threshold, namely $4.15 <W<4.25$, corresponding to the third electro-production point on the left figure. 
World data from~\cite{Camerini:1975cy,Gittelman:1975ix,Anderson:1976sd,Barate:1986fq}.
}
\label{totaljpsi}
\end{figure}

In Fig.~\ref{totaljpsi},  on the left side of the figure, we show the projected precision that is possible in experiment E12-12-006 measuring the elastic-electro and photo-production of the $J/\psi$ starting very close to threshold using the SoLID detector. On the right side of Fig.~\ref{totaljpsi} we illustrate a $\vert t-t_{min} \vert$ distribution very close to threshold where $4.15 < W < 4.25$. Here, $W$ is the invariant mass of the virtual photon nucleon system. Note that the experimental reach includes the kinematic region of the recently observed LHCb "pentaquark" when produced in the s-channel assuming a 5\% coupling following Ref.~\cite{Wang:2015jsa}. The prediction of the $t$ channel \jpsi production follows the 2 gluon exchange model of Brodsky al.~\cite{Brodsky:2000zc}. 

\subsection{A search for the LHCb pentaquark using the HMS and SHMS in Hall C}

\begin{figure}
\centering
\includegraphics[width=.33\textheight]{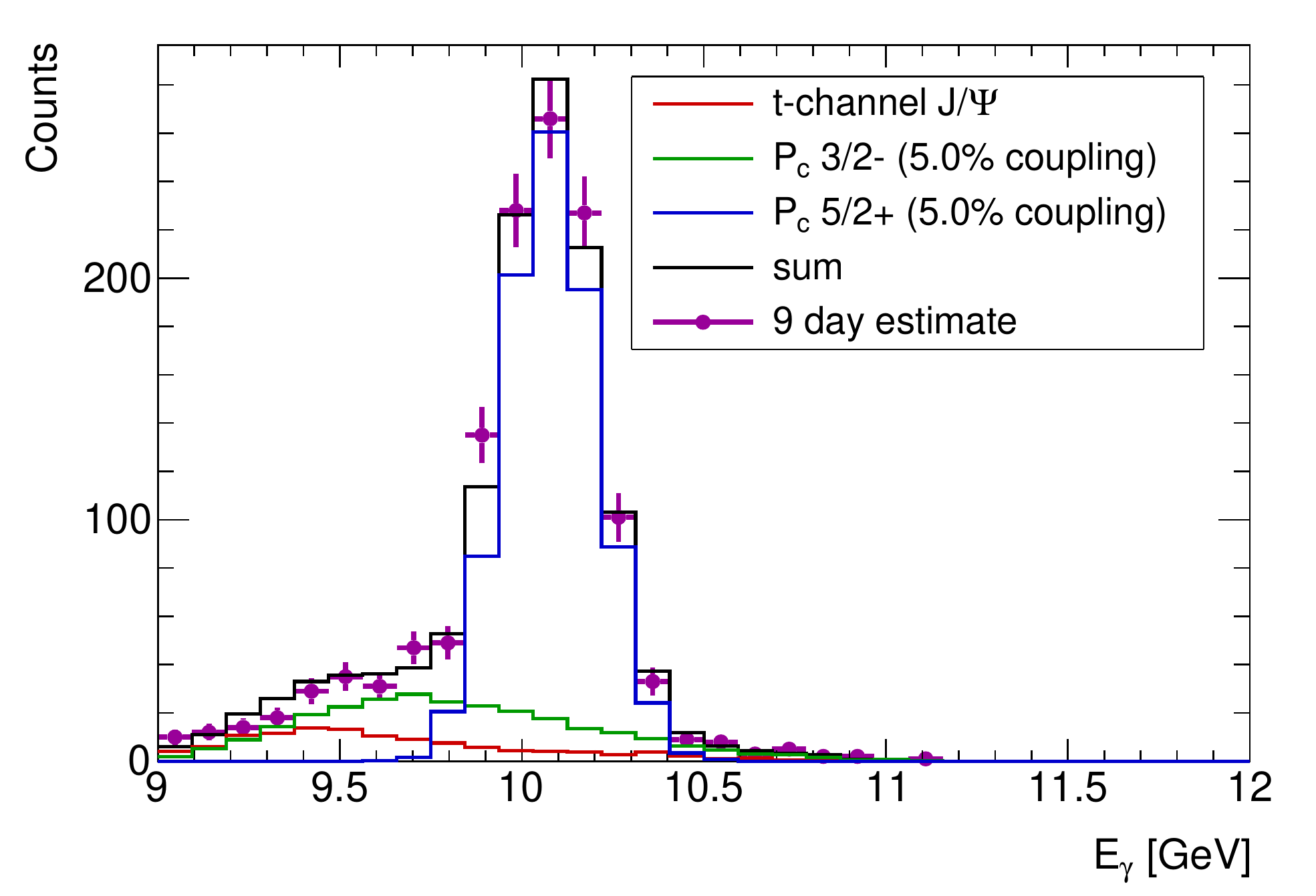}
\includegraphics[width=.33\textheight]{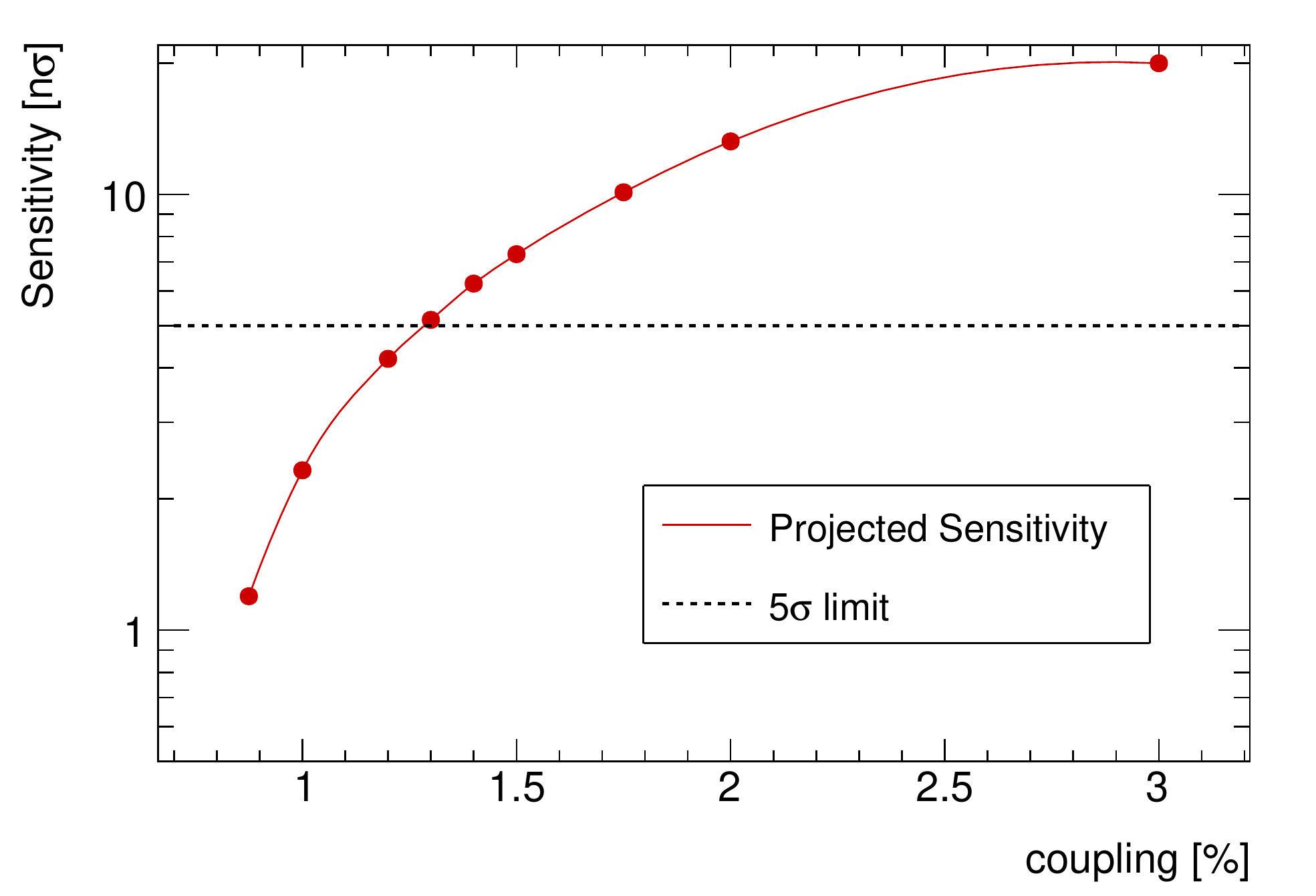}
\caption{
Left: The expected spectrum using a 5\% coupling for \pc (5/2+)-nucleon-\jpsi in a measurement of 9 days.
Right: Sensitivity to the $P_c$ as a function of the coupling to the $\jpsi p$ channel, obtained from a log-likelihood analysis.
The dashed line shows the $5\sigma$ level of sensitivity necessary for discovery.
This level is reached starting from a coupling of 1.3\%.
}
\label{figimpact}
\end{figure}

The experiment proposed and approved to search for the LHCb pentaquark in elastic photo-production in Hall C  is known as JLab E-12-16-007 and is described in~\cite{Meziani:2016lhg}. In this experiment the photo-production cross section of $J/\psi$ near threshold is measured in search of the recently observed LHCb hidden-charm
resonance $P_c$(4450) consistent with a `pentaquark'. 
A 10.6 GeV incident energy electron beam and  $50\,\mu\text{A}$ current will pass through a 9\% radiation length copper radiator to create an intense bremsstrahlung photon beam which covers the energy range of $J/\psi$ production from the threshold photo-production energy of $8.2\,\text{GeV}$, to an energy just above the presumed \pc resonance energy production.
The resulting photon beam will pass through a $15\,\text{cm}$ liquid hydrogen target,
producing \jpsi mesons through a $t$-channel diffractive process, the most common,
or through a resonant process in the $s$- and $u$-channel.
The decay $e^+e^-$ pair of the \jpsi will be detected in coincidence using the
two high-momentum spectrometers of Hall C, the high momentum spectrometer (HMS) and super high momentum spectrometer (SHMS).
The spectrometer settings have been optimized to distinguish the resonant $s$- and $u$-channel
production from the diffractive $t$-channel \jpsi production.
The $s$- and $u$-channel production of the charmed 5-quark resonance dominates the $t$-distribution at large $t$. 
The momentum and angular resolution of the spectrometers is sufficient
to observe a clear resonance enhancement in the total cross section and $t$-distribution. This is shown in Fig.~\ref{figimpact}

The observation of these resonances in photo-production will provide strong
evidence of the true nature of the LHCb states, distinguishing them
from kinematic enhancements. However, since the coupling of this state to the \jpsi nucleon system is not known it is important that the statistic of the experiment is high to set a meaningful limit on its existence or lack thereof. As shown in Fig.~\ref{figimpact} right side, at a 5$\sigma$ discovery level  we could reach a sensitivity of 1.3\% of coupling for the larger mass pentaquark in this experiment.

\section{Heavy quarkonia production at an EIC}

It has been shown that using \jpsi production on the nucleon at large values of $W$ at an EIC can provide the transverse spatial profiles of gluon distributions at different longitudinal momenta of the gluons~\cite{Accardi:2012qut}.
This will ultimately lead to the determination of the total gluon angular momentum contribution to the nucleon spin.
However, here, we argue that for consistency and in order to minimize $Q^2$ evolution corrections, a measurement of bottomium production at an EIC offers a complementary and critical probe to address both the contribution of the trace anomaly to the proton mass, as well as the total gluon angular momentum contribution to the proton spin.
In the first case, and similar to the Jefferson Lab measurement discussed above, the measurement of elastic $\Upsilon$ production on the proton close to threshold in electro-production (at small $Q^2$) or photo-production should allow us to determine the contribution of the trace anomaly to the proton mass.

In the second case, the use of $\Upsilon$ production at large $W$ should provide a more robust determination of the transverse spatial profiles of gluon densities at various momentum fractions of the gluons.The use of the heavier bottom quark (in $\upsilon$) compared to the charm quark (in \jpsi) makes all expansions in terms of inverse quark masses converges faster and provides for milder NLO corrections to extract the gluon densities from the measured data.
In the end, we have to understand both \jpsi and $\Upsilon$ production, and the gluon profiles extracted from these two processes should be universal.
Furthermore, they should also be consistent with the profiles obtained through the deep virtual Compton scattering (DVCS) measurements~\cite{Accardi:2012qut}.

\subsection{Elastic $\Upsilon$ production near threshold EIC}
\begin{figure}[h]
\centering
\includegraphics[height=.22\textheight]{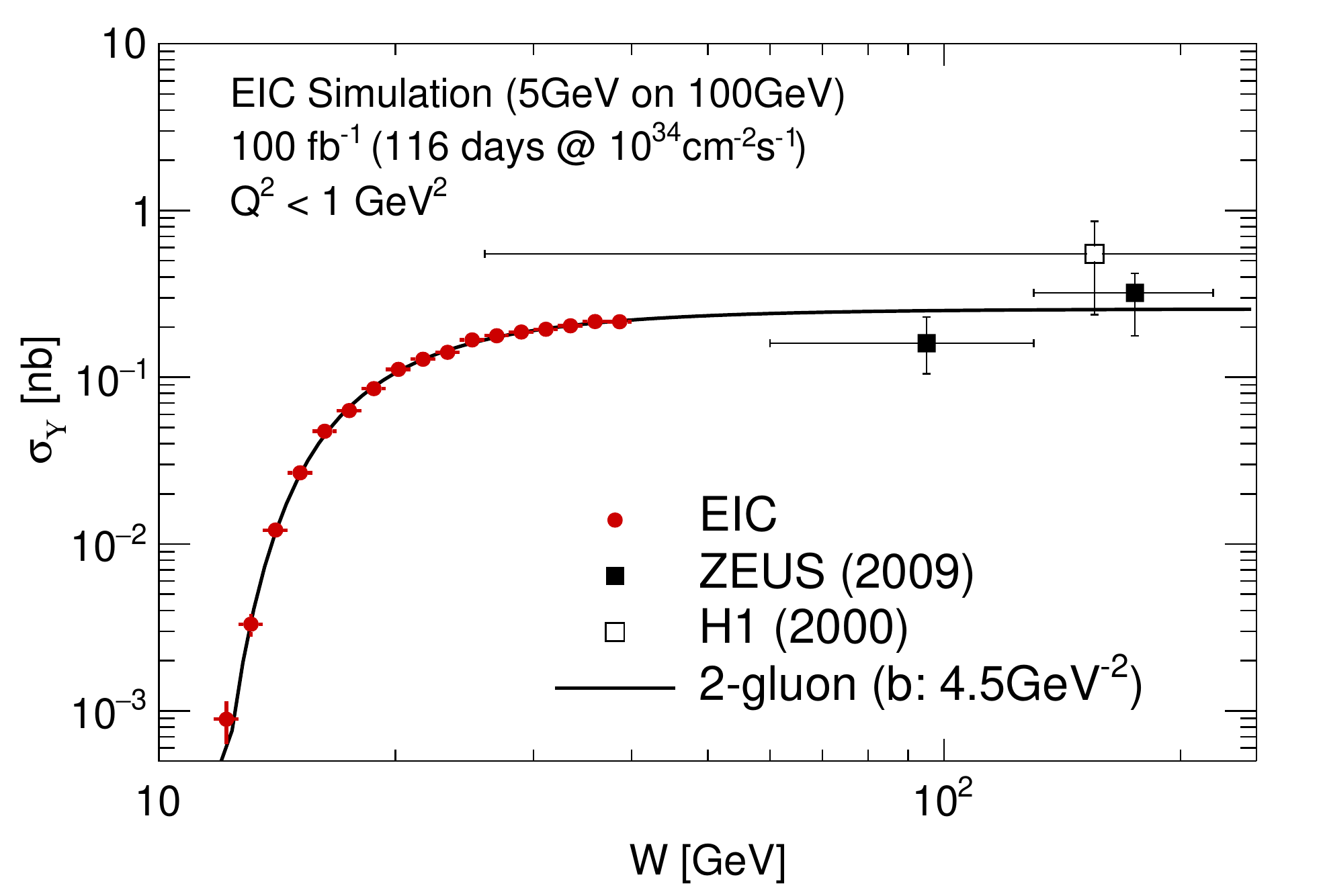}
\includegraphics[height=.22\textheight]{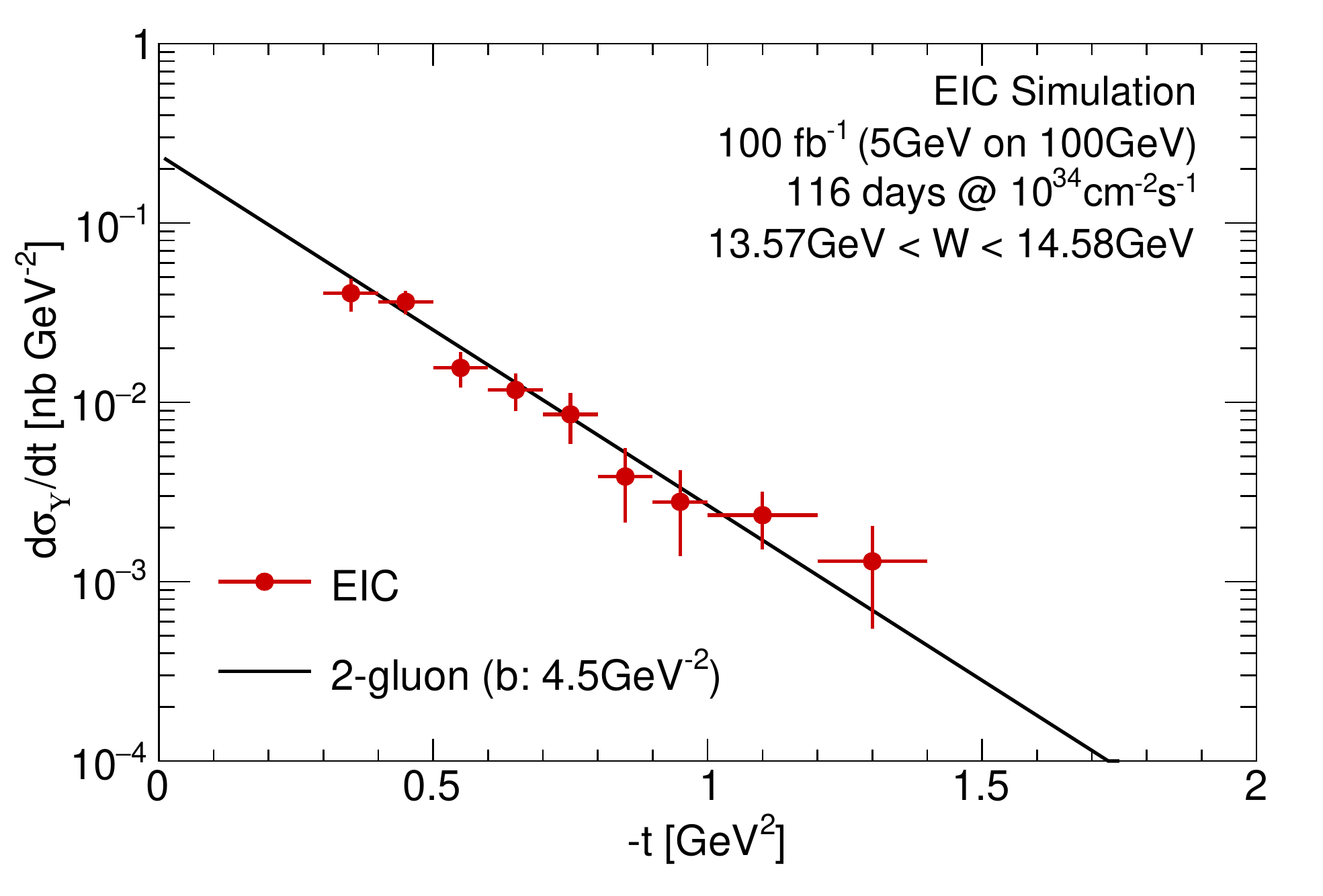}
\caption{Left: $\Upsilon$ total elastic production projections along with HERA data from the ZEUS and H1 collaborations~\cite{Chekanov:2009hv,Adloff:2000id}. Right: $t$-distribution projections for one bin interval in $W$, namely $13.57<W<14.58$ (third point on the left panel). In both cases the 2-gluon model of Brodsky et el.~\cite{Brodsky:2000zc} is used with adjusted parameters to fit the HERA data at large $W$.
\label{fig:ups-threshold}}
\end{figure}

Similar to section~\ref{jlab} but instead of using the \jpsi production we use $\Upsilon$ production to provide electro- and photo-production data very close to threshold. The projections at an EIC with a generic detector using an integrated luminosity of 100 fb$^{-1}$ are illustrated in Fig.~\ref{fig:ups-threshold} and look very promising. Using the bottomium quark would provide a much needed redundancy in the determination of the trace anomaly and would minimize the theoretical systematic uncertainties of its determination.

\subsection{Elastic $\Upsilon$ production at large $W$; Imaging the gluons}
In a study performed in the EIC white paper, it has been shown that one can access the gluonic generalized parton distribution (GPD) in the nucleons through electro-production of \jpsi at large photon-nucleon invariant mass $W$. Here we want to suggest that the electro-production of $\Upsilon$ should also be considered as part of the arsenal of an EIC to access the gluon GPD in the nucleon. The mass of the bottom quark is more then three times that of the charmed quark, and therefore most expansions in terms of the inverse of the quark masses converge faster. Next-to-leading order corrections to the leading order description of the process are typically smaller by a power of the ratio of these masses. Finally, to ensure universality, it is important to consider extracting gluon information through different processes such as DVCS, \jpsi production, and $\Upsilon$ production.   

\begin{figure}
\centering
\includegraphics[height=.2\textheight]{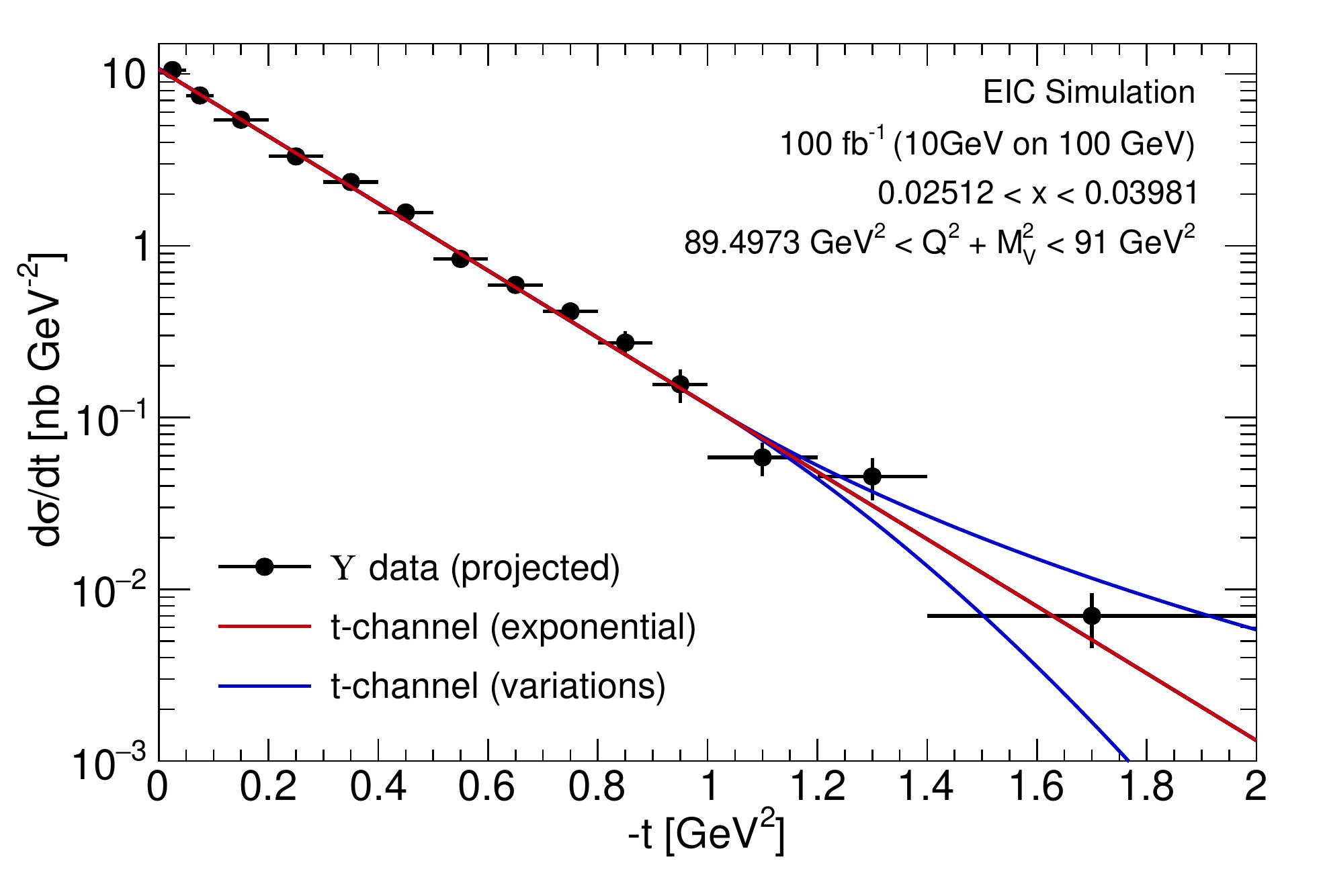}
\includegraphics[height=.2\textheight]{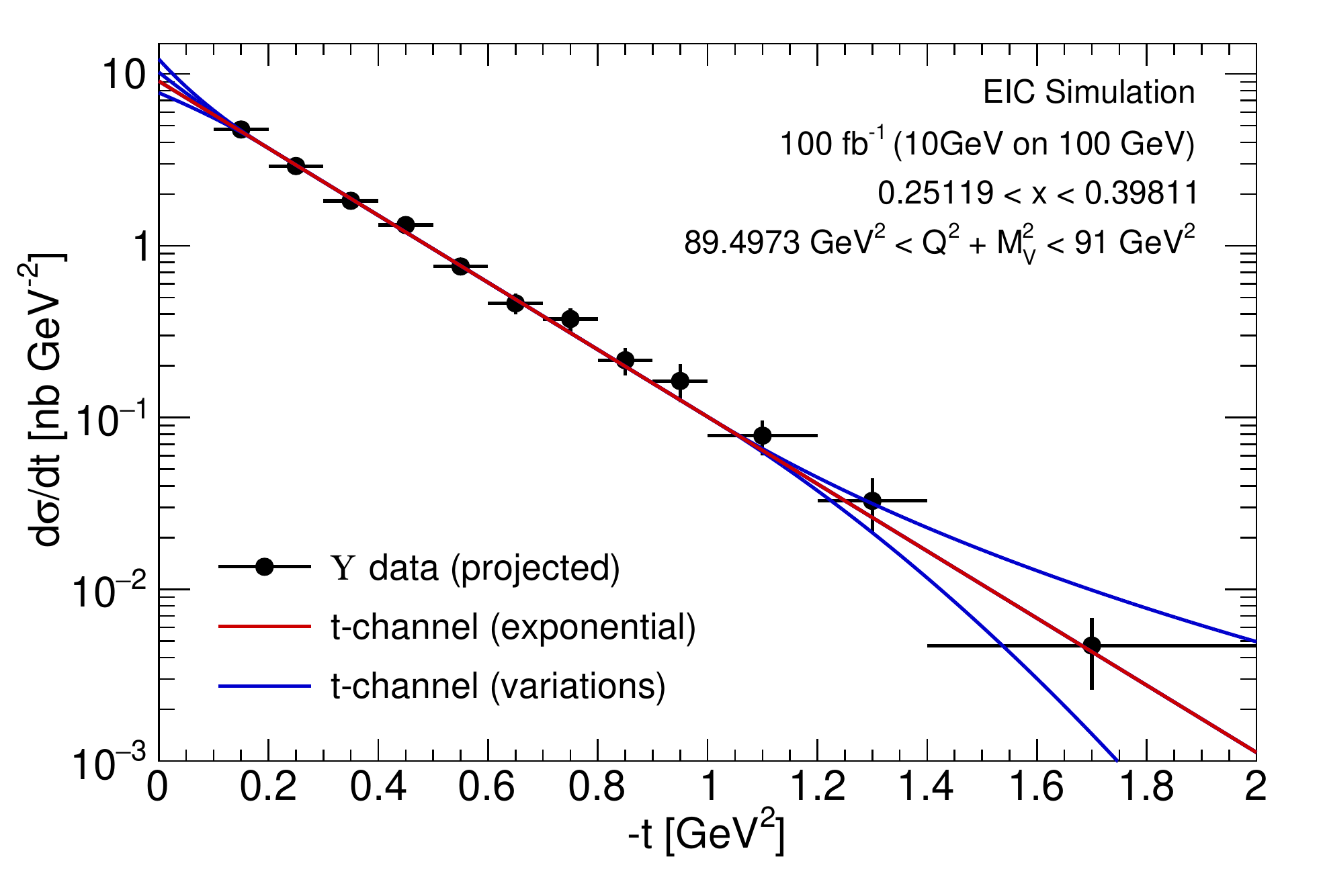}
\caption{$\Upsilon$ elastic production $t$-distributions for the lowest bin in $Q^2 + M_V^2$ at 2 different $x_V$ intervals. The red line shows the expected exponential dependence of the cross section, and the blue lines show various different extrapolations for $t$ outside of the measured region. Through Fourier transformation each of these distribution is used to provide for a spatial transverse profile of gluons.}
\label{fig:t-w-high}
\end{figure}

Fig.~\ref{fig:t-w-high} illustrates the $t$ distributions that can be achieved in a measurement at an EIC using a generic detector where both the decay of the quarkonium in $e^+e^-$ and $\mu^+\mu^-$ are accounted for. 
We assumed a generic EIC detector that can detect leptons with a pseudo-rapidity $-5 < \eta_l < 5$ and recoil protons with a polar scattering angle $\theta_p > 2\,\text{mrad}$.
Additionally, we placed a requirement of $0.01 < y < 0.8$, where $y=P\cdot q/P\cdot k$, with $P$, $k$, and $q$ the four-momenta of resp. the electron beam, the proton beam, and the virtual photon.
These requirements are consistent with the EIC white paper~\cite{Accardi:2012qut}.
We performed our simulations assuming an electron beam of 10 GeV that collides with a proton beam of 100 GeV, for an integrated luminosity of 100 fb$^{-1}$. 
This provides for a large range of possible photon-nucleon (or $\Upsilon$-nucleon) invariant masses. The left figure is a $t$ distribution in the range $0.025<x_V<0.04$ and $89.5<Q^2 + M_V^2<91$GeV$^2$ while the right figure shows the bin range $0.25<x_V<0.4$ and the same range in $Q^2 + M_V^2$. Note the factor 10 difference in $\langle x_V \rangle$ between both figures.
Here, $Q^2 + M_V^2$ is the relevant resolution scale for meson production and $x_V=(Q^2 + M_V^2)/(2P.q)$ replaces the standard Bjorken variable $x_\text{B}=Q^2/(2P\cdot q)$.

\begin{figure}
\centering
\includegraphics[width=.56\textheight]{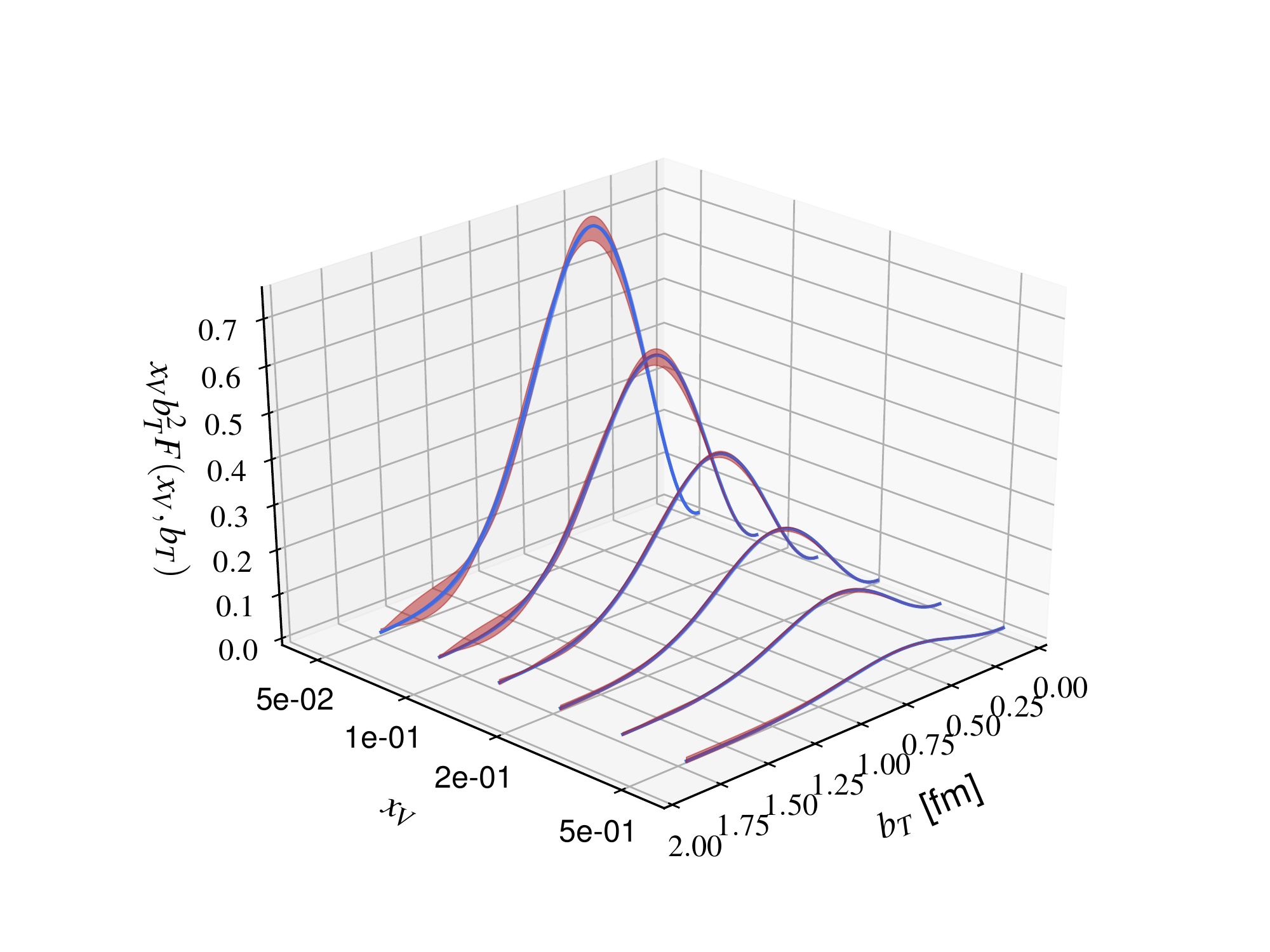}
\caption{Gluon GPD multiplied with $b_T^2$ for the lowest bin in $Q^2 + M_V^2$. This quantity highlights the precision at higher values of $b_T$. The blue band shows the statistical uncertainty of the fit, while the red band shows the statistical uncertainty added in quadrature with the systematic uncertainty due to the extrapolation at low and high $t$.
The GPDs are normalized to the gluon PDF from CT14~\cite{Dulat:2015mca}, using the LHAPDF program~\cite{Buckley:2014ana}.
}
\label{fig:profile}
\end{figure}

In Fig.~\ref{fig:profile} we show the Fourier transform of the $t$ distributions that are projected in order to obtain the spatial transverse distributions of gluons at the different $\langle x_V \rangle$ values measured. It is important to note that with the integrated luminosity chosen in this case, namely 100 fb $^{-1}$, the precision obtained on the transverse profiles is quite impressive. This measurement will complement the profiles that will be obtained from the \jpsi measurements and should be consistent after on NLO correction are performed on both measurement. It will offer the redundancy for a test of consistency and universality of the extracted gluon profiles in the nucleon.

\section{Conclusion}
In summary the mass and spin of the proton offer a laboratory to explore nucleon structure as well as our understanding of QCD. We have presented approved experiments  in Hall A and C at Jefferson Lab using the $J/\psi$  production at threshold as well as projections of possible measurements of $\Upsilon$ production at an EIC to  probe our understanding of the partition of the mass of the proton in terms of its constituents in a mass sum rule where all its pieces are accessible experimentally. We also proposed using the $\Upsilon$ production at an EIC to determine the gluon density transverse spatial profiles in a wide range of $x$ and consequently provide a path to determine the gluonic radius of the nucleon and the contribution of the total angular momentum of gluons to the nucleon spin. 

\acknowledgments
The authors thank the organizers for the opportunity to present this work. The latter is supported in part by the U.S. Department of Energy Grant Award
DE-FG02-94ER4084.

\appendix

\section{Event Generator}
We developed a new Monte-Carlo generator, LIGEN~\footnote{The source code LIGEN will be made available online.}, to
obtain a realistic estimate of the $J/\psi$ and $\Upsilon$ electro- and
photo-production rates.
LIGEN is a modular accept-reject generator written in C++14, capable of simulating
$l-A$ events for both fixed-target and collider kinematics.
Below we describe the model components used to obtain the results for this work.

\subsection{Model for the $t$-channel cross section}
In order to calculate the cross-section for the $t$-channel production, 
we fit the cross section ansatz for two-gluon exchange from Brodsky 
\textit{et al.}~\cite{Brodsky:2000zc} for
a vector meson $V$ ($J/\psi$ or $\Upsilon$),
to the available world data for $J/\psi$ and $\Upsilon$
photo-production~\cite{Camerini:1975cy,Gittelman:1975ix,Anderson:1976sd,Binkley:1981kv,Barate:1986fq,Frabetti:1993ux,Chekanov:2009hv,Adloff:2000id,Alexa:2013xxa}.

\subsection{Model for the $P_c\rightarrow J/\psi$ cross section}
Several equivalent approaches to calculate the 
$\gamma p\rightarrow P_c\rightarrow J/\psi p$ cross section can be found in the
literature~\cite{Wang:2015jsa,Karliner:2015voa,Kubarovsky:2015aaa,Blin:2016dlf}.
We based our model of the cross section on the work by Wang et al.~\cite{Wang:2015jsa}.
For more info on our approach, see Ref.~\cite{Meziani:2016lhg}

\subsection{Real photon beam}
LIGEN implements equation (24) from 
Tsai~\cite{Tsai:1966js} to evaluate the bremsstrahlung spectrum $I(E\gamma)$ when
simulating a real photon beam.

\subsection{Virtual photon spectrum and electro-production}

In case of electro-production, LIGEN simulates a virtual photon beam
from the primary electron beam using the exact relations for 
the virtual photon flux $\Gamma_T$ and polarization $\epsilon$ following the
approach from Ref.~\cite{Budnev:1975ky}.
The fully differential electro-production cross section can be parameterized as,
\begin{align}
\frac{d\sigma}{dQ^2 dy dt} &=
\Gamma_T (1 + \epsilon R) D \frac{d\sigma_V}{dt},
\end{align}
with $R=\sigma_L/\sigma_T$ following the approach from Ref.~\cite{Martynov:2003eg},
\begin{align}
R = \left(\frac{A M_V^2 + Q^2}{AM_V^2}\right)^{n_1} - 1,
\end{align}
using the parameters from Ref.~\cite{Fiore:2009ej}.
The dipole-like form factor $D$ encodes the $Q^2$ dependence of the process. 
Because $D$ is currently not known for elastic $J/\psi$ and $\Upsilon$ production, we decided to introduce
a form that was tuned to optimally describe the $Q^2$ dependence for exclusive $\rho$ production in a wide
range of kinematic regions~\cite{Airapetian:2000eo,Adamsetal:1997ft,Tytgat:2011wv,Liebing:2004uw},
\begin{align}
D = \left(\frac{M_V^2}{M_V^2 + Q^2}\right)^{n_2}.
\end{align}
Note that this form deviates from the standard dipole form factor used in a vector meson dominance model.

\subsection{Angular dependence of the decay leptons}

For the projections for the upcoming $J/\psi$ program at Jefferson Lab, we included the $J/\psi\rightarrow e^+e^-$ decay channel,
while we included the 
$\Upsilon\rightarrow e^+e^-$ and
$\Upsilon\rightarrow\mu^+\mu^-$ decay channels for the EIC-related simulations.

To describe the angular distribution of the decay products, we used the $s$-channel
helicity conservation framework (SCHC) for the case of a vector meson decaying
into two
fermions~\cite{ZEUSCollaboration:1999gb,Collaboration:2002fa,Schilling:1973en},
\begin{align}
\mathcal{W}(\cos\theta_\text{CM}) &=
\frac{3}{8}
(1 + r_{00}^{04} + (1 - 3r_{00}^{04})\cos^2\theta_\text{CM}),\\
R&=\frac{1}{\epsilon}
\frac{r_{00}^{04}}{1-r_{00}^{04}}.
\end{align}
For real photo-production, the spin-density matrix element 
$r_{00}^{04}\rightarrow 0$, simplifying
the angular dependence,
\begin{align}
\mathcal{W}(\cos\theta_\text{CM}) &=
\frac{3}{8}
(1 + \cos^2\theta_\text{CM}).
\end{align}

\bibliographystyle{JHEP}
\bibliography{references}

\providecommand{\href}[2]{#2}\begingroup\raggedright\begin{thebibliography}{10}

\bibitem{Alexandrou:2017oeh}
C.~Alexandrou, M.~Constantinou, K.~Hadjiyiannakou, K.~Jansen, C.~Kallidonis,
  G.~Koutsou et~al., \emph{{Nucleon Spin and Momentum Decomposition Using
  Lattice QCD Simulations}},
  \href{https://doi.org/10.1103/PhysRevLett.119.142002}{\emph{Phys. Rev. Lett.}
  {\bfseries 119} (2017) 142002},
  [\href{https://arxiv.org/abs/1706.02973}{{\ttfamily 1706.02973}}].

\bibitem{Dudek:2012vr}
J.~Dudek et~al., \emph{{Physics Opportunities with the 12 GeV Upgrade at
  Jefferson Lab}}, \href{https://doi.org/10.1140/epja/i2012-12187-1}{\emph{Eur.
  Phys. J.} {\bfseries A48} (2012) 187},
  [\href{https://arxiv.org/abs/1208.1244}{{\ttfamily 1208.1244}}].

\bibitem{Accardi:2012qut}
A.~Accardi et~al., \emph{{Electron Ion Collider: The Next QCD Frontier}},
  \href{https://doi.org/10.1140/epja/i2016-16268-9}{\emph{Eur. Phys. J.}
  {\bfseries A52} (2016) 268},
  [\href{https://arxiv.org/abs/1212.1701}{{\ttfamily 1212.1701}}].

\bibitem{Ji:1996ek}
X.-D. Ji, \emph{{Gauge-Invariant Decomposition of Nucleon Spin}},
  \href{https://doi.org/10.1103/PhysRevLett.78.610}{\emph{Phys. Rev. Lett.}
  {\bfseries 78} (1997) 610--613},
  [\href{https://arxiv.org/abs/hep-ph/9603249}{{\ttfamily hep-ph/9603249}}].

\bibitem{Ji:1995sv}
X.-D. Ji, \emph{{Breakup of hadron masses and energy - momentum tensor of
  QCD}}, \href{https://doi.org/10.1103/PhysRevD.52.271}{\emph{Phys. Rev.}
  {\bfseries D52} (1995) 271--281},
  [\href{https://arxiv.org/abs/hep-ph/9502213}{{\ttfamily hep-ph/9502213}}].

\bibitem{Lorce:2017xzd}
C.~Lorc\'e, \emph{{On the hadron mass decomposition}},
  \href{https://arxiv.org/abs/1706.05853}{{\ttfamily 1706.05853}}.

\bibitem{Durr:2008zz}
S.~Durr et~al., \emph{{Ab-Initio Determination of Light Hadron Masses}},
  \href{https://doi.org/10.1126/science.1163233}{\emph{Science} {\bfseries 322}
  (2008) 1224--1227}, [\href{https://arxiv.org/abs/0906.3599}{{\ttfamily
  0906.3599}}].

\bibitem{Aoki:2008sm}
{\scshape PACS-CS} collaboration, S.~Aoki et~al., \emph{{2+1 Flavor Lattice QCD
  toward the Physical Point}},
  \href{https://doi.org/10.1103/PhysRevD.79.034503}{\emph{Phys. Rev.}
  {\bfseries D79} (2009) 034503},
  [\href{https://arxiv.org/abs/0807.1661}{{\ttfamily 0807.1661}}].

\bibitem{Aoki:2009ix}
{\scshape PACS-CS} collaboration, S.~Aoki et~al., \emph{{Physical Point
  Simulation in 2+1 Flavor Lattice QCD}},
  \href{https://doi.org/10.1103/PhysRevD.81.074503}{\emph{Phys. Rev.}
  {\bfseries D81} (2010) 074503},
  [\href{https://arxiv.org/abs/0911.2561}{{\ttfamily 0911.2561}}].

\bibitem{Roberts:2016vyn}
C.~D. Roberts, \emph{{Perspective on the origin of hadron masses}},
  \href{https://doi.org/10.1007/s00601-016-1168-z}{\emph{Few Body Syst.}
  {\bfseries 58} (2017) 5}, [\href{https://arxiv.org/abs/1606.03909}{{\ttfamily
  1606.03909}}].

\bibitem{Peskin:1995ev}
M.~E. Peskin and D.~V. Schroeder, \emph{{An Introduction to quantum field
  theory}}.
\newblock Addison-Wesley, Reading, USA, 1995.

\bibitem{Kharzeev:1995ij}
D.~Kharzeev, \emph{{Quarkonium interactions in QCD}},
  \href{https://arxiv.org/abs/nucl-th/9601029}{{\ttfamily nucl-th/9601029}}.

\bibitem{Kharzeev:1998bz}
D.~Kharzeev, H.~Satz, A.~Syamtomov and G.~Zinovjev, \emph{{J / psi
  photoproduction and the gluon structure of the nucleon}},
  \href{https://doi.org/10.1007/s100529900047}{\emph{Eur.Phys.J.} {\bfseries
  C9} (1999) 459--462}, [\href{https://arxiv.org/abs/hep-ph/9901375}{{\ttfamily
  hep-ph/9901375}}].

\bibitem{Gryniuk:2016mpk}
O.~Gryniuk and M.~Vanderhaeghen, \emph{{Accessing the real part of the forward
  $J/\psi$-p scattering amplitude from $J/\psi$ photoproduction on protons
  around threshold}},
  \href{https://doi.org/10.1103/PhysRevD.94.074001}{\emph{Phys. Rev.}
  {\bfseries D94} (2016) 074001},
  [\href{https://arxiv.org/abs/1608.08205}{{\ttfamily 1608.08205}}].

\bibitem{Aaij:2015tga}
{\scshape LHCb} collaboration, R.~Aaij et~al., \emph{{Observation of $J/\psi-p$
  Resonances Consistent with Pentaquark States in $\Lambda_b^0 \to J/\psi K^-
  p$ Decays}},
  \href{https://doi.org/10.1103/PhysRevLett.115.072001}{\emph{Phys. Rev. Lett.}
  {\bfseries 115} (2015) 072001},
  [\href{https://arxiv.org/abs/1507.03414}{{\ttfamily 1507.03414}}].

\bibitem{Liu:2015fea}
X.-H. Liu, Q.~Wang and Q.~Zhao, \emph{{Understanding the newly observed heavy
  pentaquark candidates}},
  \href{https://doi.org/10.1016/j.physletb.2016.03.089}{\emph{Phys. Lett.}
  {\bfseries B757} (2016) 231--236},
  [\href{https://arxiv.org/abs/1507.05359}{{\ttfamily 1507.05359}}].

\bibitem{Karliner:2015ina}
M.~Karliner and J.~L. Rosner, \emph{{New Exotic Meson and Baryon Resonances
  from Doubly-Heavy Hadronic Molecules}},
  \href{https://doi.org/10.1103/PhysRevLett.115.122001}{\emph{Phys. Rev. Lett.}
  {\bfseries 115} (2015) 122001},
  [\href{https://arxiv.org/abs/1506.06386}{{\ttfamily 1506.06386}}].

\bibitem{Chen:2015loa}
R.~Chen, X.~Liu, X.-Q. Li and S.-L. Zhu, \emph{{Identifying exotic hidden-charm
  pentaquarks}},
  \href{https://doi.org/10.1103/PhysRevLett.115.132002}{\emph{Phys. Rev. Lett.}
  {\bfseries 115} (2015) 132002},
  [\href{https://arxiv.org/abs/1507.03704}{{\ttfamily 1507.03704}}].

\bibitem{Eides:2015dtr}
M.~I. Eides, V.~{\relax Yu}. Petrov and M.~V. Polyakov, \emph{{Narrow
  nucleon-$\psi(2S)$ bound state and LHCb pentaquarks}},
  \href{https://doi.org/10.1103/PhysRevD.93.054039}{\emph{Phys. Rev.}
  {\bfseries D93} (2016) 054039},
  [\href{https://arxiv.org/abs/1512.00426}{{\ttfamily 1512.00426}}].

\bibitem{Wang:2015jsa}
Q.~Wang, X.-H. Liu and Q.~Zhao, \emph{{Photoproduction of hidden charm
  pentaquark states $P_c^+(4380)$ and $P_c^+(4450)$}},
  \href{https://doi.org/10.1103/PhysRevD.92.034022}{\emph{Phys. Rev.}
  {\bfseries D92} (2015) 034022},
  [\href{https://arxiv.org/abs/1508.00339}{{\ttfamily 1508.00339}}].

\bibitem{Karliner:2015voa}
M.~Karliner and J.~L. Rosner, \emph{{Photoproduction of Exotic Baryon
  Resonances}},
  \href{https://doi.org/10.1016/j.physletb.2015.11.068}{\emph{Phys. Lett.}
  {\bfseries B752} (2016) 329--332},
  [\href{https://arxiv.org/abs/1508.01496}{{\ttfamily 1508.01496}}].

\bibitem{Kubarovsky:2015aaa}
V.~Kubarovsky and M.~B. Voloshin, \emph{{Formation of hidden-charm pentaquarks
  in photon-nucleon collisions}},
  \href{https://doi.org/10.1103/PhysRevD.92.031502}{\emph{Phys. Rev.}
  {\bfseries D92} (2015) 031502},
  [\href{https://arxiv.org/abs/1508.00888}{{\ttfamily 1508.00888}}].

\bibitem{Guo:2015umn}
F.-K. Guo, U.-G. Mei{\ss}ner, W.~Wang and Z.~Yang, \emph{{How to reveal the
  exotic nature of the P$_c$(4450)}},
  \href{https://doi.org/10.1103/PhysRevD.92.071502}{\emph{Phys. Rev.}
  {\bfseries D92} (2015) 071502},
  [\href{https://arxiv.org/abs/1507.04950}{{\ttfamily 1507.04950}}].

\bibitem{Blin:2016dlf}
A.~N.~H. Blin, C.~Fern\'andez-Ram\'irez, A.~Jackura, V.~Mathieu, V.~I. Mokeev,
  A.~Pilloni et~al., \emph{{Studying the P$_c$(4450) resonance in J/$\psi$
  photoproduction off protons}},
  \href{https://doi.org/10.1103/PhysRevD.94.034002}{\emph{Phys. Rev.}
  {\bfseries D94} (2016) 034002},
  [\href{https://arxiv.org/abs/1606.08912}{{\ttfamily 1606.08912}}].

\bibitem{Lu:2016nnt}
{L\"{u}, Qi-Fang and Dong, Yu-Bing}, \emph{{Strong decay mode $J/\psi p$ of
  hidden charm pentaquark states $P_c^+(4380)$ and $P_c^+(4450)$ in $\Sigma_c
  \bar{D}^*$ molecular scenario}},
  \href{https://doi.org/10.1103/PhysRevD.93.074020}{\emph{Phys. Rev.}
  {\bfseries D93} (2016) 074020},
  [\href{https://arxiv.org/abs/1603.00559}{{\ttfamily 1603.00559}}].

\bibitem{Huang:2016tcr}
Y.~Huang, J.-J. Xie, J.~He, X.~Chen and H.-F. Zhang, \emph{{Photoproduction of
  hidden-charm states in $\gamma p \to \bar{D}^{*0} \Lambda^+_c$ reaction near
  threshold}},  \href{https://arxiv.org/abs/1604.05969}{{\ttfamily
  1604.05969}}.

\bibitem{Bai:2003sw}
{\scshape BES} collaboration, J.~Z. Bai et~al., \emph{{Observation of a near
  threshold enhancement in th p anti-p mass spectrum from radiative $J /\psi$
  ---> gamma p anti-p decays}},
  \href{https://doi.org/10.1103/PhysRevLett.91.022001}{\emph{Phys. Rev. Lett.}
  {\bfseries 91} (2003) 022001},
  [\href{https://arxiv.org/abs/hep-ex/0303006}{{\ttfamily hep-ex/0303006}}].

\bibitem{Meziani:2016lhg}
Z.~E. Meziani, S.~Joosten et~al., \emph{{A Search for the LHCb Charmed
  'Pentaquark' using Photo-Production of $J/{\psi}$ at Threshold in Hall C at
  Jefferson Lab}},  \href{https://arxiv.org/abs/1609.00676}{{\ttfamily
  1609.00676}}.

\bibitem{Chen:2014psa}
{\scshape SoLID} collaboration, J.~P. Chen, H.~Gao, T.~K. Hemmick, Z.~E.
  Meziani and P.~A. Souder, \emph{{A White Paper on SoLID (Solenoidal Large
  Intensity Device)}},  \href{https://arxiv.org/abs/1409.7741}{{\ttfamily
  1409.7741}}.

\bibitem{SoLIDjpsi:proposal}
{Jefferson Lab Experiment E12-12-006, Co-spokespersons, K. Hafidi, Z.-E.
  Meziani (contact person), X. Qian, N. Sparveris and Z. Zhao, PAC40
  \url{https://www.jlab.org/exp_prog/proposals/12/PR12-12-006.pdf}}.

\bibitem{Brodsky:2000zc}
S.~J. Brodsky, E.~Chudakov, P.~Hoyer and J.~M. Laget, \emph{{Photoproduction of
  charm near threshold}},
  \href{https://doi.org/10.1016/S0370-2693(00)01373-3}{\emph{Phys. Lett.}
  {\bfseries B498} (2001) 23--28},
  [\href{https://arxiv.org/abs/hep-ph/0010343}{{\ttfamily hep-ph/0010343}}].

\bibitem{Camerini:1975cy}
U.~Camerini, J.~G. Learned, R.~Prepost, C.~M. Spencer, D.~E. Wiser, W.~Ash
  et~al., \emph{{Photoproduction of the $\psi$ Particles}},
  \href{https://doi.org/10.1103/PhysRevLett.35.483}{\emph{Phys. Rev. Lett.}
  {\bfseries 35} (1975) 483}.

\bibitem{Gittelman:1975ix}
B.~Gittelman, K.~M. Hanson, D.~Larson, E.~Loh, A.~Silverman and G.~Theodosiou,
  \emph{{Photoproduction of the psi (3100) Meson at 11-GeV}},
  \href{https://doi.org/10.1103/PhysRevLett.35.1616}{\emph{Phys. Rev. Lett.}
  {\bfseries 35} (1975) 1616}.

\bibitem{Anderson:1976sd}
R.~L. Anderson, \emph{{Excess Muons and New Results in psi Photoproduction}},
  in \emph{{International Conference on the Production of Particles with New
  Quantum Numbers Madison, Wis., April 22-24, 1976}}, p.~102, 1976,
  \href{http://www-public.slac.stanford.edu/sciDoc/docMeta.aspx?slacPubNumber=SLAC-PUB-1741}{http://www-public.slac.stanford.edu/sciDoc/docMeta.aspx?slacPubNumber=SLAC-PUB-1741}.

\bibitem{Barate:1986fq}
{\scshape NA14} collaboration, R.~Barate et~al., \emph{{Measurement of $J/\psi$
  and $\psi^\prime$ Real Photoproduction on $^{6}$Li at a Mean Energy of
  90-{GeV}}}, \href{https://doi.org/10.1007/BF01548261}{\emph{Z. Phys.}
  {\bfseries C33} (1987) 505}.

\bibitem{Chekanov:2009hv}
{\scshape ZEUS} collaboration, S.~Chekanov et~al., \emph{{Exclusive
  photoproduction of upsilon mesons at HERA}},
  \href{https://doi.org/10.1016/j.physletb.2009.07.066}{\emph{Phys. Lett.}
  {\bfseries B680} (2009) 4--12},
  [\href{https://arxiv.org/abs/0903.4205}{{\ttfamily 0903.4205}}].

\bibitem{Adloff:2000id}
{\scshape H1} collaboration, C.~Adloff et~al., \emph{{Elastic photoproduction
  of J / psi and Upsilon mesons at HERA}},
  \href{https://doi.org/10.1016/S0370-2693(00)00530-X}{\emph{Phys. Lett.}
  {\bfseries B483} (2000) 23--35},
  [\href{https://arxiv.org/abs/hep-ex/0003020}{{\ttfamily hep-ex/0003020}}].

\bibitem{Dulat:2015mca}
S.~Dulat, T.-J. Hou, J.~Gao, M.~Guzzi, J.~Huston, P.~Nadolsky et~al.,
  \emph{{New parton distribution functions from a global analysis of quantum
  chromodynamics}},
  \href{https://doi.org/10.1103/PhysRevD.93.033006}{\emph{Phys. Rev.}
  {\bfseries D93} (2016) 033006},
  [\href{https://arxiv.org/abs/1506.07443}{{\ttfamily 1506.07443}}].

\bibitem{Buckley:2014ana}
A.~Buckley, J.~Ferrando, S.~Lloyd, K.~Nordström, B.~Page, M.~Rüfenacht
  et~al., \emph{{LHAPDF6: parton density access in the LHC precision era}},
  \href{https://doi.org/10.1140/epjc/s10052-015-3318-8}{\emph{Eur. Phys. J.}
  {\bfseries C75} (2015) 132},
  [\href{https://arxiv.org/abs/1412.7420}{{\ttfamily 1412.7420}}].

\bibitem{Binkley:1981kv}
M.~E. M.~Binkley et~al., \emph{{$J/\psi$ Photoproduction from 60-GeV/c to
  300-GeV/c}}, \href{https://doi.org/10.1103/PhysRevLett.48.73}{\emph{Phys.
  Rev. Lett.} {\bfseries 48} (1982) 73}.

\bibitem{Frabetti:1993ux}
{\scshape E687} collaboration, P.~L. Frabetti et~al., \emph{{A Measurement of
  elastic $J/\psi$ photoproduction cross-section at Fermilab E687}},
  \href{https://doi.org/10.1016/0370-2693(93)90679-C}{\emph{Phys. Lett.}
  {\bfseries B316} (1993) 197--206}.

\bibitem{Alexa:2013xxa}
{\scshape H1} collaboration, C.~Alexa et~al., \emph{{Elastic and
  Proton-Dissociative Photoproduction of J/psi Mesons at HERA}},
  \href{https://doi.org/10.1140/epjc/s10052-013-2466-y}{\emph{Eur. Phys. J.}
  {\bfseries C73} (2013) 2466},
  [\href{https://arxiv.org/abs/1304.5162}{{\ttfamily 1304.5162}}].

\bibitem{Tsai:1966js}
Y.-S. Tsai and V.~Whitis, \emph{{Thick target bremsstrahlung and target
  consideration for secondary particle production by electrons}},
  \href{https://doi.org/10.1103/PhysRev.149.1248}{\emph{Phys. Rev.} {\bfseries
  149} (1966) 1248--1257}.

\bibitem{Budnev:1975ky}
V.~M. Budnev, I.~F. Ginzburg, G.~V. Meledin and V.~G. Serbo, \emph{{The Two
  photon particle production mechanism. Physical problems. Applications.
  Equivalent photon approximation}},
  \href{https://doi.org/10.1016/0370-1573(75)90009-5}{\emph{Phys. Rept.}
  {\bfseries 15} (1975) 181--281}.

\bibitem{Martynov:2003eg}
E.~Martynov, E.~Predazzi and A.~Prokudin, \emph{{Photoproduction of vector
  mesons in the soft dipole pomeron model}},
  \href{https://doi.org/10.1103/PhysRevD.67.074023}{\emph{Phys. Rev.}
  {\bfseries D67} (2003) 074023},
  [\href{https://arxiv.org/abs/hep-ph/0207272}{{\ttfamily hep-ph/0207272}}].

\bibitem{Fiore:2009ej}
R.~Fiore, L.~L. Jenkovszky, V.~K. Magas, S.~Melis and A.~Prokudin,
  \emph{{Exclusive J/Psi electroproduction in a dual model}},
  \href{https://doi.org/10.1103/PhysRevD.80.116001}{\emph{Phys. Rev.}
  {\bfseries D80} (2009) 116001},
  [\href{https://arxiv.org/abs/0911.2094}{{\ttfamily 0911.2094}}].

\bibitem{Airapetian:2000eo}
{\scshape HERMES} collaboration, A.~Airapetian et~al., \emph{{Exclusive
  leptoproduction of rho0 mesons from hydrogen at intermediate virtual photon
  energies}}, \href{https://doi.org/10.1007/s100520000483}{\emph{Eur. Phys. J.}
  {\bfseries C17} (2000) 389--398},
  [\href{https://arxiv.org/abs/hep-ex/0004023}{{\ttfamily hep-ex/0004023}}].

\bibitem{Adamsetal:1997ft}
{\scshape E665} collaboration, M.~R. Adams et~al., \emph{{Diffractive
  production of $\rho^0(770)$ mesons in muon proton interactions at 470-GeV}},
  \href{https://doi.org/10.1007/s002880050386}{\emph{Z. Phys.} {\bfseries C74}
  (1997) 237--261}.

\bibitem{Tytgat:2011wv}
M.~Tytgat, \emph{{Diffractive Production of $\rho^0$ and $\omega$ Vector Mesons
  at HERMES}}, Ph.D. thesis, Ghent University, 2011.

\bibitem{Liebing:2004uw}
P.~Liebing, \emph{{Can the gluon polarization in the nucleon be extracted from
  HERMES data on single high-$p_T$ hadrons.}}, Ph.D. thesis, University of
  Hamburg, 2004.

\bibitem{ZEUSCollaboration:1999gb}
{\scshape ZEUS} collaboration, J.~Breitweg et~al., \emph{{Exclusive
  electroproduction of $\rho^0$ and $J/\psi$ mesons at HERA}},
  \href{https://doi.org/10.1007/s100529901051}{\emph{Eur. Phys. J.} {\bfseries
  C6} (1999) 603--627}, [\href{https://arxiv.org/abs/hep-ex/9808020}{{\ttfamily
  hep-ex/9808020}}].

\bibitem{Collaboration:2002fa}
{\scshape ZEUS} collaboration, S.~Chekanov et~al., \emph{{Exclusive
  photoproduction of J / psi mesons at HERA}},
  \href{https://doi.org/10.1007/s10052-002-0953-7}{\emph{Eur. Phys. J.}
  {\bfseries C24} (2002) 345--360},
  [\href{https://arxiv.org/abs/hep-ex/0201043}{{\ttfamily hep-ex/0201043}}].

\bibitem{Schilling:1973en}
K.~Schilling and G.~Wolf, \emph{{How to analyze vector meson production in
  inelastic lepton scattering}},
  \href{https://doi.org/10.1016/0550-3213(73)90371-4}{\emph{Nucl. Phys.}
  {\bfseries B61} (1973) 381--413}.

\end{thebibliography}\endgroup

\end{document}